\documentclass{PoS}
\pdfoutput=1

\newcommand{\BR}{{\cal B}}
\newcommand{\lambdacp}{\Lambda_{c}^{+}}
\newcommand{\lambdacm}{\overline{\Lambda}{}_{c}^{-}}
\newcommand{\modea}{pK_{S}^0}
\newcommand{\modeb}{pK^{-}\pi^+}
\newcommand{\modec}{pK_{S}^0\pi^0}
\newcommand{\moded}{pK_{S}^0\pi^+\pi^-}
\newcommand{\modee}{pK^{-}\pi^+\pi^0}
\newcommand{\modeaa}{\Lambda\pi^+}
\newcommand{\modebb}{\Lambda\pi^+\pi^0}
\newcommand{\modedd}{\Lambda\pi^+\pi^-\pi^+}
\newcommand{\modeaaa}{\Sigma^{0}\pi^+}
\newcommand{\modeccc}{\Sigma^{+}\pi^0}
\newcommand{\modeddd}{\Sigma^{+}\pi^+\pi^-}
\newcommand{\modeeee}{\Sigma^{+}\omega}

\def\Journal#1#2#3#4{{#1} {\bf #2}, #3 (#4)}
\def\PRL{Phys. Rev. Lett.}
\def\PRB{Phys. Rev. B}
\def\PRD{Phys. Rev. D}
\def\PLB{Phys. Lett. B}

\title{Charmed hadron decays at BESIII}

\ShortTitle{Charmed hadron decays at BESIII}

\author{\speaker{Liaoyuan Dong}\thanks{on behalf of the BESIII collaboration. This work is supported in part by National Natural Science Foundation of China (NSFC) under Contracts Nos. 11075174, 11475185.}\\
        Institute of High Energy Physics, Beijing 100049, China\\
        E-mail: \email{dongly@ihep.ac.cn}}

\abstract{I present here a selection of preliminary results on charmed hadron decays from BESIII
collaboration, including the study of $D^{+}\to K^{-}\pi^{+}e^{+}\nu_{e}$, the measurement of the form factors in $D^{+}\to \omega e^{+}\nu_{e}$ and the search for $D^{+}\to \phi e^{+}\nu_{e}$, the study of decay dynamics and \emph{CP} asymmetry in $D^{+}\to K^{0}_{L} e^{+}\nu_{e}$, and the measurements of the absolute branching fractions of twelve Cabbibo-favored hadronic $\lambdacp$ decay modes and $\Lambda^+_c\rightarrow \Lambda e^+\nu_e$. The results are based on the data samples collected with the BESIII detector at the $\psi(3770)$ peak and at central-of-mass energy of $4.599$~GeV.}

\FullConference{The European Physical Society Conference on High Energy Physics\\
		22--29 July 2015\\
		Vienna, Austria}

\begin{document}

\section{Introduction}
Recent results from the BESIII experiment based on 2.92 ${\rm fb}^{-1}$ recorded at the $\psi(3770)$ peak and
567 ${\rm pb}^{-1}$ at $E_{cm} = 4.599$ GeV are presented here for studies of the charmed hadron decays.
The $\psi(3770)$ predominantly decays to pairs of $D$ mesons, either $D^+D^-$ or $D^0\bar{D}^0$.
At $E_{cm} = 4.599$ GeV, $\Lambda_c$ mesons are primarily produced as  $\lambdacp\lambdacm$ pairs.
To identify the $D\bar{D}$/$\lambdacp\lambdacm$ pairs, we make use of the double-tag technique initially
used by MARK III~\cite{doubletag}. In this technique, the yields of single tags (ST), where one $D$/$\Lambda_c$
is reconstructed in the tag modes, and double tags (DT), where both $D$/$\Lambda_c$ mesons are reconstructed,
are determined. In this report, $D^-$ is reconstructed in one of the six tag modes: $D^- \to K^+ \pi^- \pi^-$, $K^+ \pi^- \pi^- \pi^0$, $K_S^0 \pi^-$, $K_S^0 \pi^- \pi^0$, $K_S^0 \pi^- \pi^- \pi^+$, and $K^+ K^- \pi^-$, while $\lambdacp$ is reconstructed in one of twelve tag modes:
$\lambdacp \to \modea$, $\modeb$, $\modec$, $\moded$, $\modee$, $\modeaa$, $\modebb$, $\modedd$, $\modeaaa$, $\modeccc$, $\modeddd$ and $\modeeee$.
The hadronic decays are identified using the beam-constrained mass $M_{BC}\equiv \sqrt{E_{\rm beam}^2-p^2c^2}$, where $E_{\rm beam}$ is the beam energy and  $p$ is measured momentum of $D$ or $\Lambda_c$.
The semileptonic decays are detected through the kinematic variable $U_{\rm miss}\equiv E_{\rm miss}-c|\vec{p}_{\rm miss}|$, where $E_{miss}$ and $P_{miss}$ are the missing energy and momentum carried by the neutrino, respectively.
Throughout this report, the inclusion of charge conjugated processes is implied, unless otherwise explicitly mentioned.

In this proceeding, I report five preliminary measurements from the BESIII collaboration.
First I will present three results about $D$ semileptonic decays, then present the measurements of absolute branching fractions for the twelve hadronic $\lambdacp$ decay modes,
and end this report with the measurement of absolute branching fraction of $\Lambda^+_c\rightarrow \Lambda e^+\nu_e$.

\section{$D^{+}\to K^{-}\pi^{+}e^{+}\nu_{e}$ (preliminary)}
The semileptonic decay $D^{+}\to K^{-}\pi^{+}e^{+}\nu_{e}$ provides a unique tool
for investigating the $K\pi$ system,  and measuring the $\bar{K}^*(892)^0$ resonance parameters and the hadronic transition form factors.

Using the double-tag technique, we select 18262 candidate events and measure the branching fractions to be
$\BR (D^+ \to K^{-}\pi^{+}e^{+}\nu_{e}) = (3.71\pm0.03\pm0.08)$\% over the full $m_{K\pi}$ range
and $\BR (D^+ \to K^{-}\pi^{+}e^{+}\nu_{e})_{[0.8,1]} = (3.33\pm0.03\pm0.07)$\% in the $\overline{K}^{*}(892)^{0}$ region, respectively.

\begin{figure}[htb]
  \begin{center}
    \includegraphics[width=0.80\textwidth]{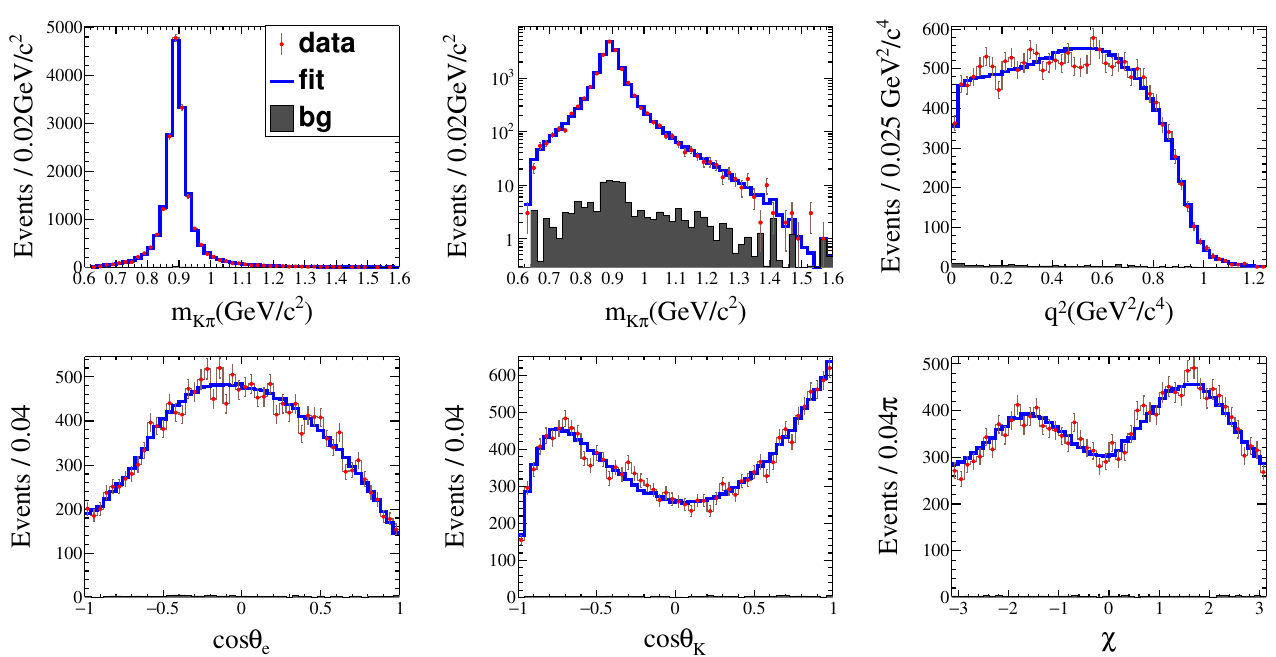}
    \caption{Projections of data (dots with error bars) and of the PWA solution (line) onto each of the kinematic variables.
             The shadowed histogram is the estimated background.}
    \label{fig1}
  \end{center}
\end{figure}

We perform a partial wave analysis (PWA) on the selected candidates.
The probability density function (PDF) is expressed on the five kinematic variables~\cite{pwa}:
$m^{2}$ ($K\pi$ mass square), $q^{2}$ ($e\nu_{e}$ mass square),
$\theta_{K}$ (angle between the $\pi$ and the $D$ direction in the $K\pi$ rest frame),
$\theta_{e}$ (angle between the $\nu_{e}$ and the $D$
direction in the $e\nu_{e}$ rest frame),
and $\chi$ (angle between the two decay planes).
The $q^{2}$ dependent helicity basis form factors are parameterized according to the spectroscopic pole dominance (SPD) model,
the phase $\delta_{S}$ of $S$-wave amplitude is parameterized as that used in the LASS parameterization.
The PWA shows that the dominant component is $\bar K^{*}(892)^{0}$,
S-wave contribution equal to $(6.05\pm0.22\pm0.18)$\%, and contributions from $\bar K^{*}(1410)^{0}$ and $\bar K^{*}_2(1430)^{0}$ are negligible.
Projections of the five kinematic variables are shown in Fig.~\ref{fig1}.
We determine the $\bar K^{*}(892)^{0}$ resonance parameters:
$m_{\bar K^{*}(892)^{0}}=(894.60\pm0.25\pm0.08)~ \rm MeV/c^2$,
$\Gamma_{\bar K^{*}(892)^{0}}=(46.42\pm0.56\pm0.15)~ \rm MeV/c^2$,
and the Blatt-Weisskopf parameter
$r_{BW}=3.07\pm0.26\pm0.11 ~(\rm GeV/c)^{-1}$.
We also measure the parameters defining the hadronic form factors:
$r_{V} = \frac{V(0)}{A_{1}(0)} = 1.411\pm0.058\pm0.007$,
$r_{2} = \frac{A_{2}(0)}{A_{1}(0)} = 0.788\pm0.042\pm0.008$,
$m_{V} = (1.81^{+0.25}_{-0.17}\pm0.02)~ \rm MeV/c^{2}$ (first measurement),
$m_{A} = (2.61^{+0.22}_{-0.17}\pm0.03)~ \rm MeV/c^{2}$, and
$A_{1}(0) = 0.585\pm0.011\pm0.017$.

\begin{figure}[htb]
  \centering
  \includegraphics[width=0.40\linewidth]{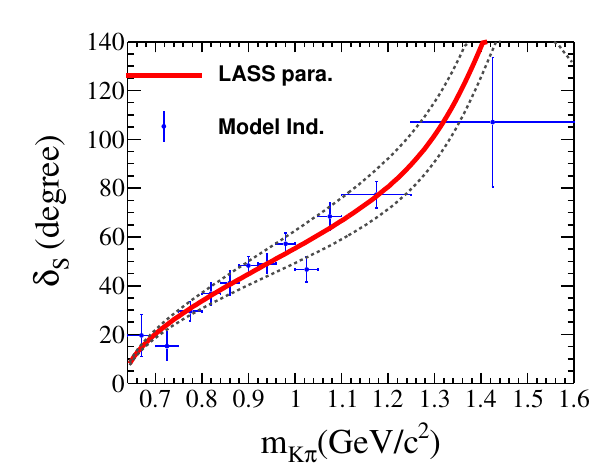}
  \caption{The \emph S-wave phase variation versus $m_{K\pi}$.
  The points with error bars are model-independent measurements,
  the solid line (the dotted line shows the one sigma deviation from the central line) is the PWA solution based on LASS parameterization.}
  \label{fig2}
\end{figure}

To test the applicability of the LASS parameterization for the S-wave phase,
we measure the phase variation of the S-wave in a model-independent way.
We divide the $m_{K\pi}$ spectrum into twelve bins
and perform the PWA fit to measure $\delta_{S}$
in each bin (the phase is assumed to be constant within each bin).
Figure~\ref{fig2} shows the comparison of
the model-independent measurement with the results
based on the LASS parameterization.
We find good agreement between both determinations of the S-wave's phase variation.

\begin{figure}[htb]
  \centering
  \includegraphics[width=0.80\linewidth]{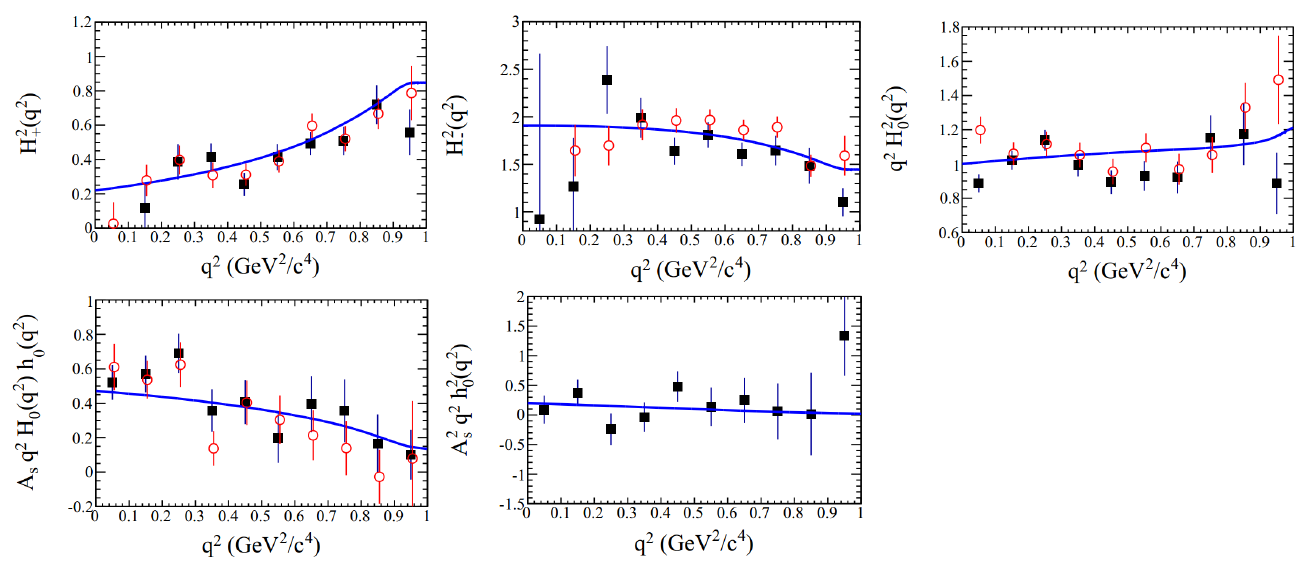}
  \caption{Form factors measured in a model-independent way (squares) compared with the CLEO-c results (circles) and with the PWA solution based on SPD model (curves).}
  \label{fig3}
\end{figure}

In the $\bar{K}^{*}(892)^{0}$ region, the decay intensity of $D^{+}\to K^{-}\pi^{+}e^{+}\nu_{e}$
can be parameterized by three helicity basis form factors:
$H_{+}(q^2,m)$, $H_{-}(q^2,m)$ and $H_{0}(q^2,m)$, describing
the decay into the vector meson $\bar{K}^{*}(892)^{0}$,
and by an additional form factor $h_{0}(q^2,m)$ describing
the non-resonant $S$-wave contribution.
We extract the helicity basis form factors in a model-independent way using
the projective weighting technique~\cite{Link},
The results are shown in Fig.~\ref{fig3}.
We find the model-independent measurements are consistent with the SPD model
with our PWA solution. Our measurements are also consistent with the CLEO-c results~\cite{Briere}.

\section{$D^{+}\to \omega e^{+}\nu_{e}$ and $D^{+}\to \phi e^{+}\nu_{e}$ (preliminary)}
\label{sec:omegaphienu}

The decay $D^{+}\to \omega e^{+}\nu_{e}$ has been observed at the CLEO-c experiment~\cite{Dobbs}.
Neglecting lepton mass, the transition rate for $D^{+}\to \omega e^{+}\nu_{e}$ decays
depends on three dominant form factors: two axial and one vector, $A_1$, $A_2$ and $V$.
The decay $D^{+}\to \phi e^{+}\nu_{e}$ has not been observed~\cite{Yelton}.
Its rate relative to $D_s^{+}\to \omega e^{+}\nu_{e}$
will provide information about $\omega-\phi$ mixing, as well as about the non-perturbative ``weak annihilation''.

\begin{figure}[htb]
  \centering
  \includegraphics[width=0.80\linewidth]{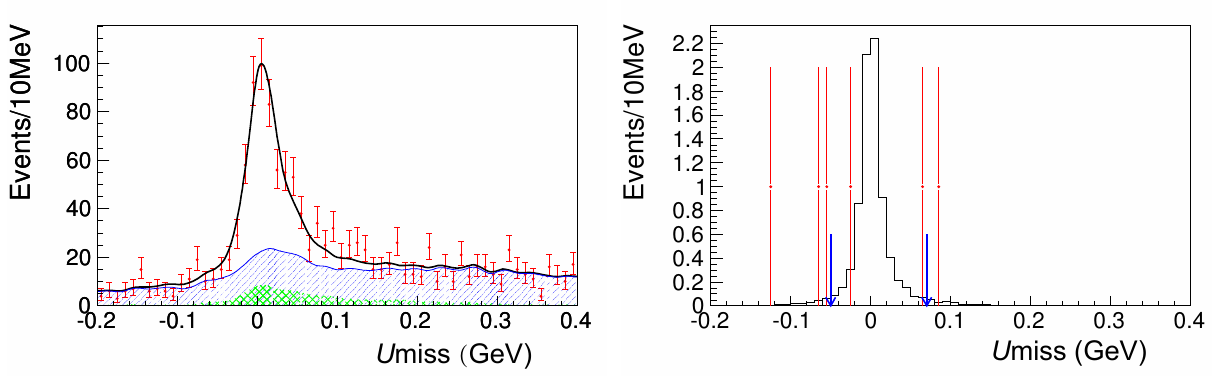}
  \caption{Left: fit to the $U_{\rm miss}$ distribution for $D^{+}\to \omega e^{+}\nu_{e}$.
  Right: the $U_{\rm miss}$ distribution for $D^{+}\to \phi e^{+}\nu_{e}$.
  The points with error bars are data.
  In left plot, the solid line is the fit,
  the filled curve is the total background, and the cross-hatched curve is the peaking background.
  In right plot, the solid histogram is the signal MC with arbitrary normalization, and the arrows show the signal region.}
  \label{fig4}
\end{figure}

With the double-tag technique, the $U_{\rm miss}$ distributions with all tag modes combined for $D^{+}\to \omega e^{+}\nu_{e}$ and $D^{+}\to \phi e^{+}\nu_{e}$ are shown in Fig.~\ref{fig4}.
For the decay $D^{+}\to \omega e^{+}\nu_{e}$, the signal yield is obtained from the fit to the $U_{\rm miss}$ distribution.
For the decay $D^{+}\to \phi e^{+}\nu_{e}$, we observed 2 events in the signal region ([-0.05, 0.07] GeV) with $4.2\pm1.5$ background.
The absolute branching fraction of the decay  $D^{+}\to \omega e^{+}\nu_{e}$ and the upper limit on the $\BR(D^{+}\to \phi e^{+}\nu_{e})$
at 90\% C.L. are listed in Table~\ref{tab1}. These results are the most precise measurements to date.

\begin{table}[htb]
\small
\begin{center}
\caption{Measured branching fractions and a comparison with the previous measurements.
For $\BR(D^{+}\to \omega e^{+}\nu_{e})$, the first uncertainty
is statistical and the second systematic.}
\begin{tabular}{lcc}
\hline\hline
Mode &  Measured $\BR$ &  Previous~\cite{Dobbs,Yelton}  \\ \hline
$\omega e^{+}\nu_{e}$  &   $(1.63\pm0.11\pm0.08)\times 10^{-3}$
    &   $(1.82\pm0.18\pm0.07)\times 10^{-3}$    \\
$\phi e^{+}\nu_{e}$    &  $1.3\times10^{-5}$ (90\%C.L.)
  &     $9.0\times10^{-5}$ (90\%C.L.) \\
\hline\hline
\end{tabular}
\label{tab1}
\end{center}
\end{table}

\begin{figure}[htb]
  \centering
  \includegraphics[width=0.80\linewidth]{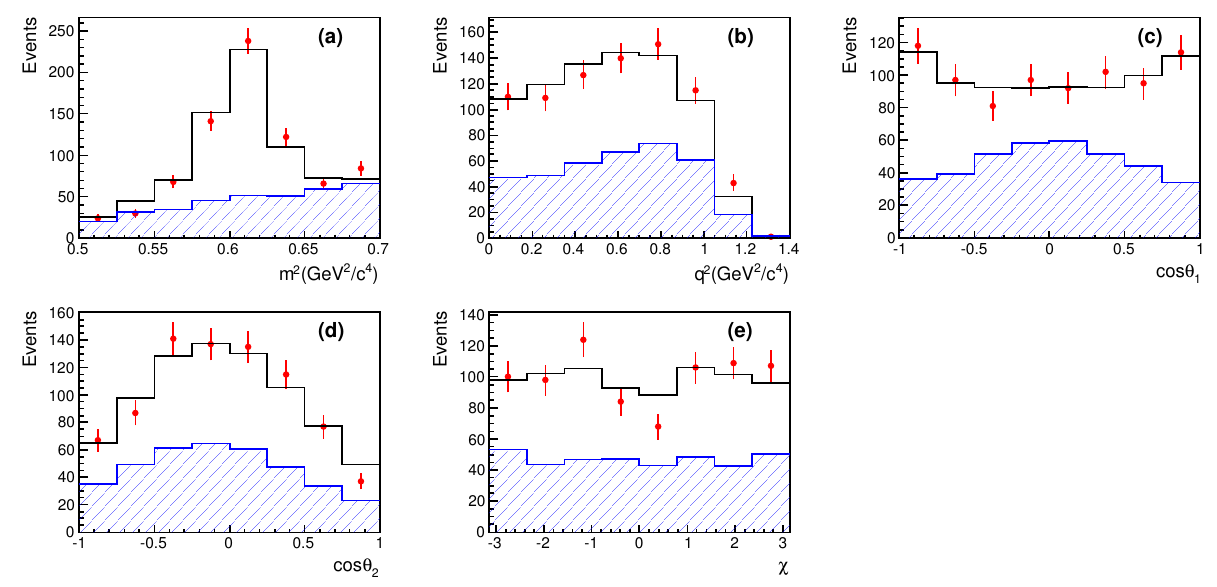}
  \caption{Projections of the data (points with error bars), and of the fit
  (solid histograms) onto each of the kinematic variables. The filled histogram curves show the
  background distributions.}
  \label{fig5}
\end{figure}

We perform a five-dimensional maximum likelihood fit in the space of
$m^2$ (mass square of $\pi\pi\pi$, $q^2$,  $\cos\theta_{1}$ (helicity angle of $\omega$), $\cos\theta_{2}$ (helicity angle of $e$) and $\chi$ (angle between the decay plans),
to measure the form factors in the decay $D^{+}\to \omega e^{+}\nu_{e}$.
The form factor ratios are determined from the fit:
$r_{V}=\frac{V(0)}{A_1(0)}=1.24\pm0.09\pm0.06$, $r_{2}=\frac{A_2(0)}{A_1(0)}=1.06\pm0.15\pm0.05$,
which are measured for the first time.
Projections of the five kinematic variables are shown in Fig.~\ref{fig5}.

\section{$D^{+}\to K^{0}_{L} e^{+}\nu_{e}$ (preliminary)}
\label{sec:kenu}

We present the first measurement of the absolute branching fraction and the $CP$ violation for the
decay $D^{+}\to K^{0}_{L} e^{+}\nu_{e}$. With the double-tag technique,
we measure the branching fractions of the six tag modes separately for $D^{+}$ and $D^{-}$. We obtain
$\mathcal{B}(D^{+}\to K^{0}_{L} e^{+}\nu_{e}) = (4.454\pm0.038\pm0.102)\%$,
$\mathcal{B}(D^{-}\to K^{0}_{L} e^{-}\bar{\nu}_{e}) = (4.507\pm0.038\pm0.104)\%$,
which are the weighted averages of the six tag modes.
We obtain the averaged branching fraction
$\mathcal{B}(D^{+}\to K^{0}_{L} e^{+}\nu_{e}) = (4.481\pm0.027\pm0.103)$\%,
which agrees well with the measurement of $\mathcal{B}(D^{+}\to K^{0}_{S} e^{+}\nu_{e})$ by CLEO-c~\cite{Besson}.
We also obtain the $CP$ asymmetry
$A_{CP} \equiv \frac{\mathcal{B}(D^{+}\to K^{0}_{L} e^{+}\nu_{e})-\mathcal{B}(D^{-}\to K^{0}_{L} e^{-}\bar{\nu}_{e})}
{\mathcal{B}(D^{+}\to K^{0}_{L} e^{+}\nu_{e})+\mathcal{B}(D^{-}\to K^{0}_{L} e^{-}\bar{\nu}_{e})}
=(-0.59\pm0.60\pm1.48)$\%,
which is consistent with the theoretical prediction in Ref.~\cite{Xing}.

In the limit of zero electron mass, the differential decay rate for $D^{+}\to K^{0}_{L} e^{+}\nu_{e}$
depends only on one form factor $f_{+}(q^2)$.
We perform simultaneous fits to the distributions of the observed
candidates as a function of $q^2$ for six tag modes to determined the $f_{+}^{K}(0)|V_{cs}|$.
We use several parameterizations for the $f_{+}(q^2)$: the simple pole model,
the modified pole model, two-parameter series expansion, and three-parameter series expansion.
Figure~\ref{fig6} shows the simultaneous fits using the two-parameter series expansion model,
corresponding to the results: $f_{+}^{K}(0)|V_{cs}|=0.728\pm0.006\pm0.011$, and shape parameter $r_{1}=-1.91\pm0.33\pm0.24$.

\begin{figure}[htb]
  \centering
  \includegraphics[width=0.70\linewidth]{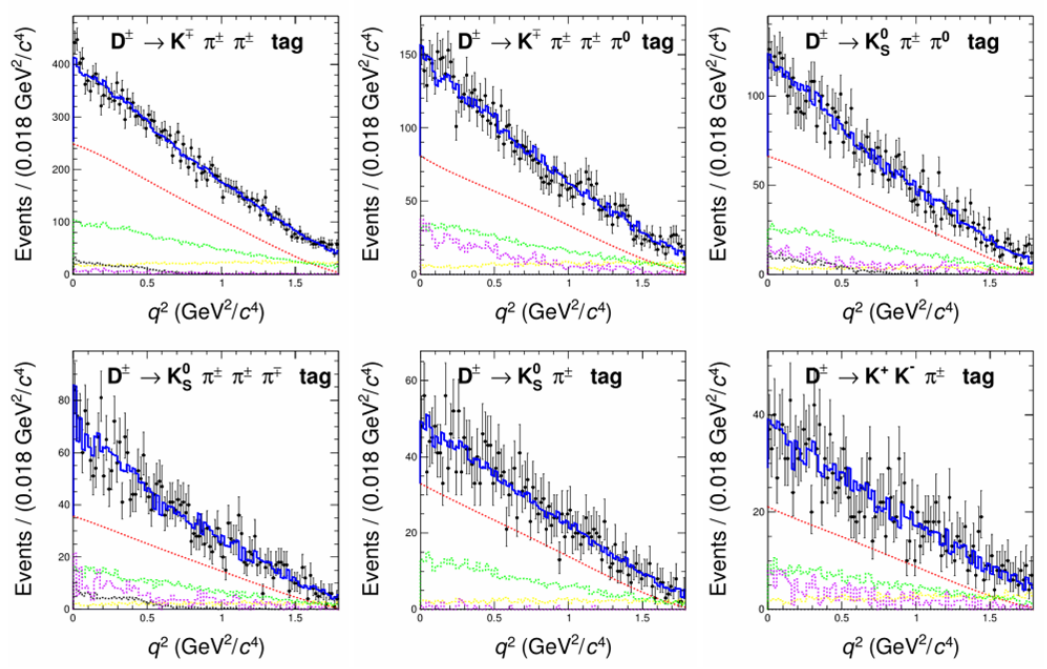}
  \caption{Fits (blue solid curves)
  to observed candidates (points with error bars) as a function of $q^2$ by two-parameter series expansion parameterization.
  In each plot, the red dashed curves show the signal, while the yellow, violet, black and green curves refer to different kinds of backgrounds.}
  \label{fig6}
\end{figure}

\section{ $\Lambda_c^+$ Hadronic decays (preliminary)}
\label{sec:lambda-had}
Hadronic $\Lambda_c^+$ decays rates are key probes to understand $b$-flavor meson and baryon decays.
Experimentally, most of the branching fractions of the $\Lambda_c^+$ are measured referring to the $pK^-\pi^+$ mode.
However, PDG has made a model-dependent determination of the absolute $\BR(\Lambda_c^+\to pK^-\pi^+)=(5.0\pm1.3)$\% with large uncertainty~\cite{2014pdg}.
Recently, Belle reports a model-independent measurement of $\BR(\Lambda_c^+\to pK^-\pi^+)=(6.84\pm0.24^{+0.21}_{-0.27})$\%, which improve the precision by a factor of 5.

Using the double-tag technique, we measure the absolute branching fractions for twelve Cabbibo-favored hadronic $\Lambda_c^+$ decay modes.
The ST and DT yields are obtained by fitting the $M_{BC}$ distributions of the $\Lambda_c$ candidates.
We perform a least square fit, which considers statistical and systematic correlations among different hadronic modes,
to obtain the branching fractions of the twelve $\lambdacp$ decay modes globally.
Table~\ref{tab2} lists the resultant ST yields and DT yields, as well as the fitted branching fractions of $\lambdacp$, where the uncertainties are statistical only.
Our result on $\BR(\Lambda_c^+\to pK^-\pi^+)$ is consistent with that in PDG, but lower than Belle's with a significance of about $2\sigma$.
For the branching fractions of the other modes,  the precisions of our measurement are significantly improved  comparing to the world average values in PDG.

\begin{table}[htb]
\small
\begin{center}
\caption{ST yields, DT yields, and the measured branching fractions with comparison to PDG values and Belle measurement.
For our results, the uncertainties are statistical only. The branching fractions do not include any sub decay rates.}
\begin{tabular}{lccccc}
\hline\hline
  Mode                & ST yield    & DT yield    & Measured $\BR$ (\%) & PDG $\BR$ (\%) & Belle $\BR$(\%)~\cite{Zupanc}\\ \hline
$\textbf{$\modea$}$   & $1243\pm37$ & $89\pm10$   & $1.48\pm0.08$ & $1.15\pm0.30$ & \\
$\textbf{$\modeb$}$   & $6308\pm88$ & $390\pm21$  & $5.77\pm0.27$ & $5.0\pm1.3$   & $6.84\pm0.24^{+0.21}_{-0.27}$\\
$\textbf{$\modec$}$   & $558\pm33$  & $40\pm7$    & $1.77\pm0.12$ & $1.65\pm0.50$ & \\
$\textbf{$\moded$}$   & $454\pm28$  & $29\pm6$    & $1.43\pm0.10$ & $1.30\pm0.35$ & \\
$\textbf{$\modee$}$   & $1849\pm71$ & $148\pm14$  & $4.25\pm0.22$ & $3.4\pm1.0$   & \\
$\textbf{$\modeaa$}$  & $706\pm27$  & $59\pm8$    & $1.20\pm0.07$ & $1.07\pm0.28$ & \\
$\textbf{$\modebb$}$  & $1497\pm52$ & $89\pm11$   & $6.70\pm0.35$ & $3.6\pm1.3$   & \\
$\textbf{$\modedd$}$  & $609\pm31$  & $53\pm7$    & $3.67\pm0.23$ & $2.6\pm0.7$   & \\
$\textbf{$\modeaaa$}$ & $586\pm32$  & $39\pm6$    & $1.28\pm0.08$ & $1.05\pm0.28$ & \\
$\textbf{$\modeccc$}$ & $271\pm25$  & $20\pm5$    & $1.18\pm0.11$ & $1.00\pm0.34$ & \\
$\textbf{$\modeddd$}$ & $836\pm43$  & $56\pm8$    & $3.58\pm0.22$ & $3.6\pm1.0$   & \\
$\textbf{$\modeeee$}$ & $157\pm22$  & $13\pm3$    & $1.47\pm0.18$ & $2.7\pm1.0$   & \\
\hline\hline
\end{tabular}
\label{tab2}
\end{center}
\end{table}

\section{$\Lambda^+_c\rightarrow \Lambda e^+\nu_e$ (preliminary)}
The $\Lambda^+_c\rightarrow \Lambda e^+\nu_e$ decay provide a good test on non-perturbation theoretical models and calibrate the calculations of lattice quantum chromodynamics (LQCD) in charm baryon sector.

Using the similar strategy in Section~\ref{sec:lambda-had}, we obtain the signal yield by fitting the $U_{\rm miss}$ distribution for the candidate events, as shown in Fig.~\ref{fig6}.
We obtain the number of the signals to be $103.5\pm10.9$ after subtracting the number of background.
The absolute branching fraction for $\Lambda_c^+\rightarrow \Lambda e^+\nu_e$ is determined to be $\mathcal B({\Lambda^+_c\rightarrow \Lambda e^+\nu_e})=(3.63\pm0.38)\%$,
where the error is statistical only. Our result improves the precision of PDG value more than twofold.
\begin{figure}[htb]
  \centering
  \includegraphics[width=0.45\linewidth]{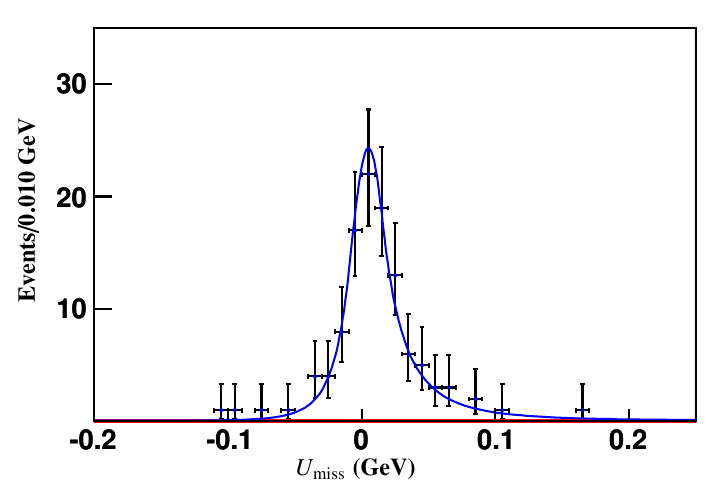}
  \caption{Fit to the $U_{\rm miss}$ distribution for $\Lambda_c^+\rightarrow \Lambda e^+\nu_e$.
  The points with error bars are data, the (blue) solid line shows the total fit and the (red) dashed line is the background shape.}
  \label{fig6}
\end{figure}

\end{document}